\documentclass[epj]{webofc}
\usepackage[varg]{txfonts}   
\newcommand{\be}{\begin{equation}}
\newcommand{\ee}{\end{equation}}
\newcommand{\bea}{\begin{eqnarray}}
\newcommand{\eea}{\end{eqnarray}}
\wocname{New Frontiers in Physics}
\woctitle{New Frontiers in Physics 2014}
\begin{document}
\title{Parton/hadron dynamics in heavy-ion collisions at FAIR energies}

\author{W. Cassing\inst{1}\fnsep\thanks{\email{Wolfgang.Cassing@theo.physik.uni-giessen.de}} \and
        V. P. Konchakovski\inst{1}\and
        A. Palmese\inst{1} \and
        V. D. Toneev\inst{2}
 \and
        E. L. Bratkovskaya\inst{3}
}

\institute{Institut f\"ur Theoretische Physik, University of
Giessen, Germany
\and
           Joint Institute for Nuclear Research,
  141980 Dubna, Russia
\and
           Institut f\"ur Theoretische Physik,
 Johann Wolfgang Goethe University Frankfurt/M,
 Germany
          }

\abstract{ Recent STAR data for the directed flow of protons,
antiprotons and charged pions obtained within the beam energy scan
program are analyzed within the Parton-Hadron-String-Dynamics
(PHSD/HSD) transport models. Both versions of the kinetic approach
are used to clarify the role of partonic degrees of freedom. The
PHSD results, simulating a partonic phase and its coexistence with a
hadronic one, are roughly consistent with the STAR data. Generally,
the semi-qualitative agreement between the measured data and model
results supports the idea  of a crossover type of quark-hadron
transition which softens the nuclear EoS but shows no indication of
a first-order phase transition. Furthermore, the directed flow of
kaons and antikaons is evaluated in the PHSD/HSD approachesfrom
$\sqrt{s_{NN}} \approx$ 3 - 200 GeV which shows a high sensitivity
to hadronic potentials in the FAIR/NICA energy regime $\sqrt{s_{NN}}
\le$ 8 GeV.}
\maketitle
\section{Introduction}
\label{intro} The study of  the particle azimuthal angular
distribution in momentum space with respect to the reaction plane is
an important tool to probe the hot, dense matter created in
heavy-ion collisions \cite{VPS10,So10}. The directed flow refers to
a collective sidewards deflection of particles and is characterized
by the first-order harmonic $v_1$ of the Fourier expansion of the
particle azimuthal angular distribution with respect to the reaction
plane~\cite{PV98}. The second harmonic coefficient $v_2$, called
elliptic flow, and the triangular flow $v_3$ have been extensively
studied both theoretically and experimentally in the last years by
about five orders of magnitude in the collision energy
$\sqrt{s_{NN}}$~\cite{STAR10}. In contrast, apart from first
measurements in the early nineties and till recent times, the
directed flow  was studied mainly theoretically
\cite{Danil,Sahu,JEM} although some experimental information from
the Schwerionen-Synchrotron (SIS) to Super-Proton-Synchrotron (SPS)
energies is available \cite{CBMbook}.

It is generally assumed that  the directed flow is  generated during
the nuclear passage time~\cite{So97,HWW99}. The directed transverse
flow therefore probes the onset of bulk collective dynamics during
thermalization, thus providing valuable information on the
pre-equilibrium stage~\cite{SH92,KKP95,E877,NA44}.  In earlier times
(at moderate beam energies) the first flow harmonic defined as
\begin{equation} v_1(y)=\left< \cos(\phi-\phi_{RP})\right>=\left<
v_x/\sqrt{v_x^2+v_y^2}\right> \end{equation} with respect to the
reaction plane $\phi_{RP}$ was characterized differently: i.e. by
the mean transverse momentum per particle projected on the reaction
($x-z$) plane $\left<p_x(y)/N\right>$ in the center-of-mass system
which differs from the $v_1$ harmonic component.  Unfortunately, it
is not possible to convert or directly compare $v_1$ data to the
earlier $p_x/N$ analysis. The NA49 collaboration~\cite{NA49} has
measured the flow coefficient v$_1$ for pions and protons at SPS
energies and a negative $v_1(y)$ slope was observed by the standard
event plane method for pions. Often, just the slope of $v_1(y)$ at
midrapidity has been used to quantify the strength of the directed
flow.

At Alternating-Gradient-Synchrontron (AGS) energies $E_{lab}\lesssim
$11.5 A$\cdot$GeV, the $v_1$ dependence has a characteristic S-shape
attributed to the standard $\left<p_x(y)/N\right>$ distribution. The
projected average momentum $<p_x(y)>$ grows linearly with rising
rapidity $y$ between the target and projectile fragmentation
regions. Conventionally, this type of flow- with positive derivative
$dv_1/dy$ - is called normal flow, in contrast to the antiflow for
which $dv_1/dy <$0~\cite{E877-7,NA49,RR97,HWW99}. At these  moderate
energies the slope of $v_1(y)$ at midrapidity ($F$) is observed to
be positive for protons and significantly smaller in magnitude and
negative for pions~\cite{E877-7,NA49,WA98}. The smooth fall-off of
this function with  beam energy is reasonably reproduced by the
available hadronic kinetic models (see the comparison in
Ref.~\cite{INN00}).

The shape of the  rapidity dependence $v_1(y)$ with bombarding
energy is of special interest because the directed flow at
midrapidity may be modified by the collective expansion and reveal a
signature of a phase transition from normal nuclear matter to a
quark-gluon plasma (QGP). This is commonly studied by measuring the
central rapidity region that reflects important  features of the
system evolution from its initial state. The predicted $v_1(y)$ flow
coefficient is small close to midrapidity with almost no dependence
on pseudorapidity. However, as first demonstrated in
Refs.~\cite{Ri95, Ri96}, the 3D hydrodynamic expansion -  with an
equation of state (EoS) including a possible phase transition -
exhibits some irregularity in the evolution of the system. When
including a first order phase transition  this leads to a local
minimum in the proton excitation function of the transverse directed
flow at $E_{lab}\approx$8 A$\cdot$GeV. Such a first order transition
leads to a softening of the equation of state and consequently to a
time-delayed expansion. The existence of this 'softest point' of {
the EoS} at a minimum of the energy density $\varepsilon_{SP}$ leads
to a long lifetime of the mixed phase and consequently in a
prolonged expansion of matter~\cite{HS94}. Presently, the critical
energy density (or latent heat for a first order transition at
finite quark chemical potential) is not well known and estimates
vary from 0.5 GeV/fm$^3$ to 1.5
GeV/fm$^3$~\cite{HS94,ST95,MO95,RG96}. A softest point at
$\varepsilon_{SP}\sim$1.5 GeV/fm$^3$ should give a minimum in the
directed flow excitation function at $E_{lab}\sim$ 30 GeV
A$\cdot$GeV~\cite{HS94,ST95}. In case of ideal hydrodynamics the
directed proton flow $p_x$ shows even a negative $v_1$ ('$v_1$
collapse') between $E_{lab}=$8 and 20 A$\cdot$GeV~\cite{St05} and
with rising energy increases back to a positive flow.  The ideal
hydro calculations suggest that this 'softest point collapse' is at
$E_{lab}\sim $8 A$\cdot$GeV but this was not confirmed by available
AGS data~\cite{St05}. A linear extrapolation of the AGS data
indicates that a collapse of the directed proton flow might be at
$E_{lab}\approx $30 A$\cdot$GeV. However, this minimum in the given
energy range is not supported in the two-fluid model with a phase
transition~\cite{INN00}.

The interest in the directed flow $v_1(y)$  has recently been
enhanced considerably due to new STAR data obtained in the framework
of the beam energy scan (BES) program~\cite{STAR-14}. The directed
flow of identified hadrons -- protons, antiprotons, positive and
negative pions -- has been measured with high precision for
semi-central Au+Au collisions in the energy range
$\sqrt{s_{NN}}=$(7.7-39) GeV. These data provide a promising basis
for studying direct-flow issues as discussed above and have been
addressed already by the Frankfurt group~\cite{SAP14} limiting
themselves to the energy $\sqrt{s_{NN}}<$20 GeV where hadronic
processes are expected to be dominant. However, the authors of
Ref.~\cite{SAP14} did not succeed to describe the data and to obtain
conclusive results which led to the notion of the 'directed flow
puzzle'. Our study aims to analyze these STAR results in the whole
available energy range including in particular antiproton data
\cite{Volodya2014}.

We start with a short reminder of the PHSD approach and its hadronic
version HSD (without partonic degrees of freedom) and then analyse
the BES data in terms of both transport models in order to explore
where effects from partonic degrees of freedom show up. Furthermore,
we compare also with predictions of other kinetic models in Sec. II.
In Sec. III we  provide predictions for kaon and antikaon directed
flows for Au+Au collisions from $\sqrt{s_{NN}} \approx$ 3 - 200 GeV
and investigate in particular the sensitivity to hadronic
potentials. Our findings are summarized in Sec. IV.

\section{Directed flow in microscopic approaches}
\label{Sec2}
\subsection{Reminder of PHSD}
\label{phsdd}
 The PHSD model is a covariant dynamical approach for
strongly interacting systems formulated on the basis of
Kadanoff-Baym equations~\cite{JCG04,CB09} or off-shell transport
equations in phase-space representation, respectively. In the
Kadanoff-Baym theory the field quanta are described in terms of
dressed propagators with complex selfenergies. Whereas the real part
of the selfenergies can be related to mean-field potentials of
Lorentz scalar, vector or tensor type, the imaginary parts provide
information about the lifetime and/or reaction rates of time-like
particles~\cite{Ca09}. Once the proper complex selfenergies of the
degrees of freedom are known, the time evolution of the system is
fully governed by off-shell transport equations for quarks and
hadrons (as described in Refs.~\cite{JCG04,Ca09}). The PHSD model
includes the creation of massive quarks via  hadronic string decay -
above the critical energy density $\sim 0.5$ GeV/fm$^3$ - and quark
fusion forming a hadron in the hadronization process. With some
caution, the latter process can be considered as a simulation of a
crossover transition since the underlying EoS in PHSD is a
crossover~\cite{Ca09}. At energy densities close to the critical
energy density the PHSD describes a coexistence of the quark-hadron
mixture. This approach allows for a simple and transparent
interpretation of lattice QCD results for thermodynamic quantities
as well as correlators and leads to effective strongly interacting
partonic quasiparticles with broad spectral functions. For a review
on off-shell transport theory we refer the reader to
Ref.~\cite{Ca09}; PHSD model results and their comparison with
experimental observables for heavy-ion collisions from the lower
super-proton-synchrotron (SPS) to RHIC energies can be found in
Refs.~\cite{To12,KBC12,Linnyk2011,Ca09}. In the hadronic phase, i.e.
for energies densities below the critical energy density, the PHSD
approach is identical to the Hadron-String-Dynamics (HSD)
model~\cite{EC96,PhysRep,CBJ00}.

\begin{figure}[thb]
\centering
\sidecaption
\includegraphics[width=8cm,clip]{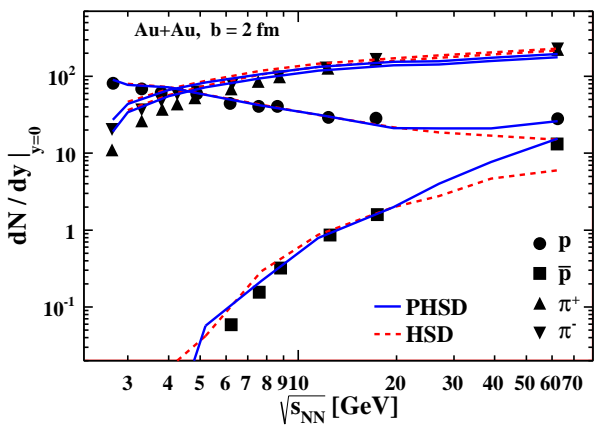}
\caption{Particle abundance at mid-rapidity calculated for central
collisions $b=$2 fm in the HSD (dashed lines) and PHSD (solid lines)
models. The experimental data are from a compilation of
Ref.~\cite{ABS06} complemented by recent data from the STAR
collaboration \cite{Zhu12} and the latest update of the compilation
of NA49 results~\cite{BGL,BM11}. } \label{fig1}
\end{figure}

The  HSD approach  formally can be written as a coupled set of
transport equations for the phase-space distributions $f_h(x,p)$ of
hadron $h$, which includes the real part of the scalar and vector
hadron self-energies. The hadron quasi-particle properties here are
defined via the mass-shell constraint with effective masses and
momenta. In the HSD transport calculations we include nucleons,
$\Delta$'s, $N^*(1440), N^\star(1535), \Lambda, \Sigma$ and
$\Sigma^\star$ hyperons, $\Xi$'s and $\Omega$'s as well as their
antiparticles. High energy inelastic hadron-hadron collisions are
described by the FRITIOF model { \cite{Fritiof}}, where two
incoming hadrons emerge the reaction as two excited color singlet
states, i.e. 'strings'. The excitation functions for various
dynamical quantities as well as experimental observables from SIS to
RHIC energies within the HSD transport approach can be found in
Refs.~\cite{PhysRep,CBJ00,Brat04}.

Fig.~\ref{fig1} illustrates how the hadron multiplicity $dN/dy(y=0)$
at midrapidity is reproduced within the PHSD (solid lines) and HSD
(dashed lines) kinetic approaches. We point out that the antiproton
abundance is a crucial issue. In the AGS-SPS low energy range ($\le$
20 GeV) both models agree quite reasonably with experiment,
including the antiproton yield. The enhancement of the proton and
antiproton yield at $\sqrt{s_{NN}}$ = 62 GeV in PHSD relative to HSD
can be traced back to a larger baryon/antibaryon fraction in the
hadronization process. At lower energies this agreement is reached
by taking into account the $p\bar p$ annihilation to three mesons
(e.g. $\pi, \rho, \omega$) as well as the inverse channels employing
detailed balance as worked out in Ref.~\cite{Ca02}. These inverse
channels are quite important; in particular, at the top SPS energy
this inverse reaction practically compensates the loss of
antiprotons due to their annihilation~\cite{Ca02}. At lower SPS and
AGS energies the annihilation is dominant due to the lower meson
abundancies, however, the backward channels reduce the net
annihilation rate. We mention that the multiple-meson recombination
channels are not incorporated in the standard UrQMD transport
model~\cite{Bass}. The proton multiplicities are reproduced rather
well in the PHSD/HSD approaches but the multiplicity of charged
pions is slightly overestimated for $\sqrt{s_{NN}}\le$10 GeV. This
discrepancy is observed also in other transport
models~\cite{Br99,LCLM01}.

\subsection{Directed flow from microscopic transport models}
\label{transport}

\begin{figure}[thb]
\includegraphics[width=0.495\textwidth]{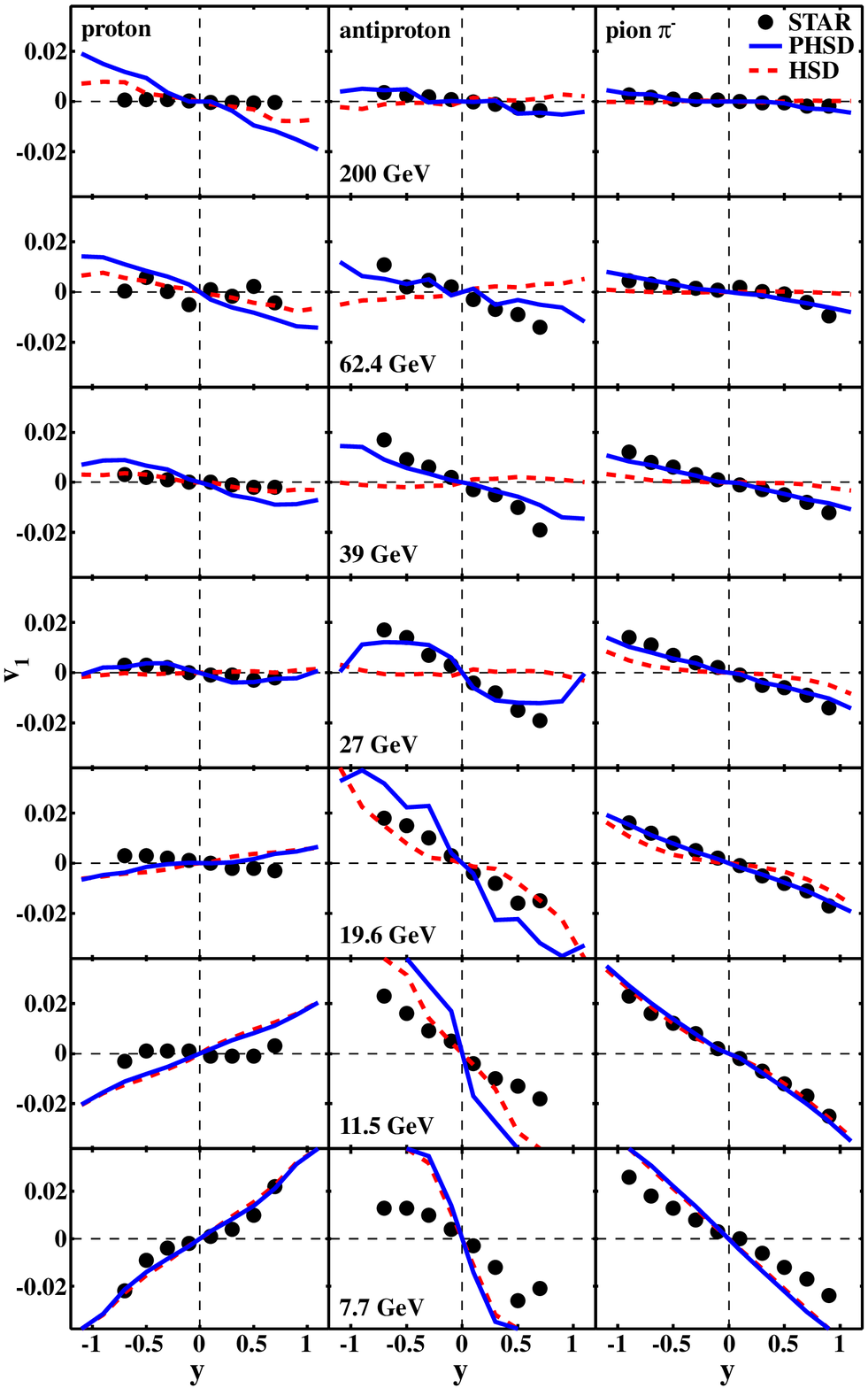} \includegraphics[width=0.48\textwidth]{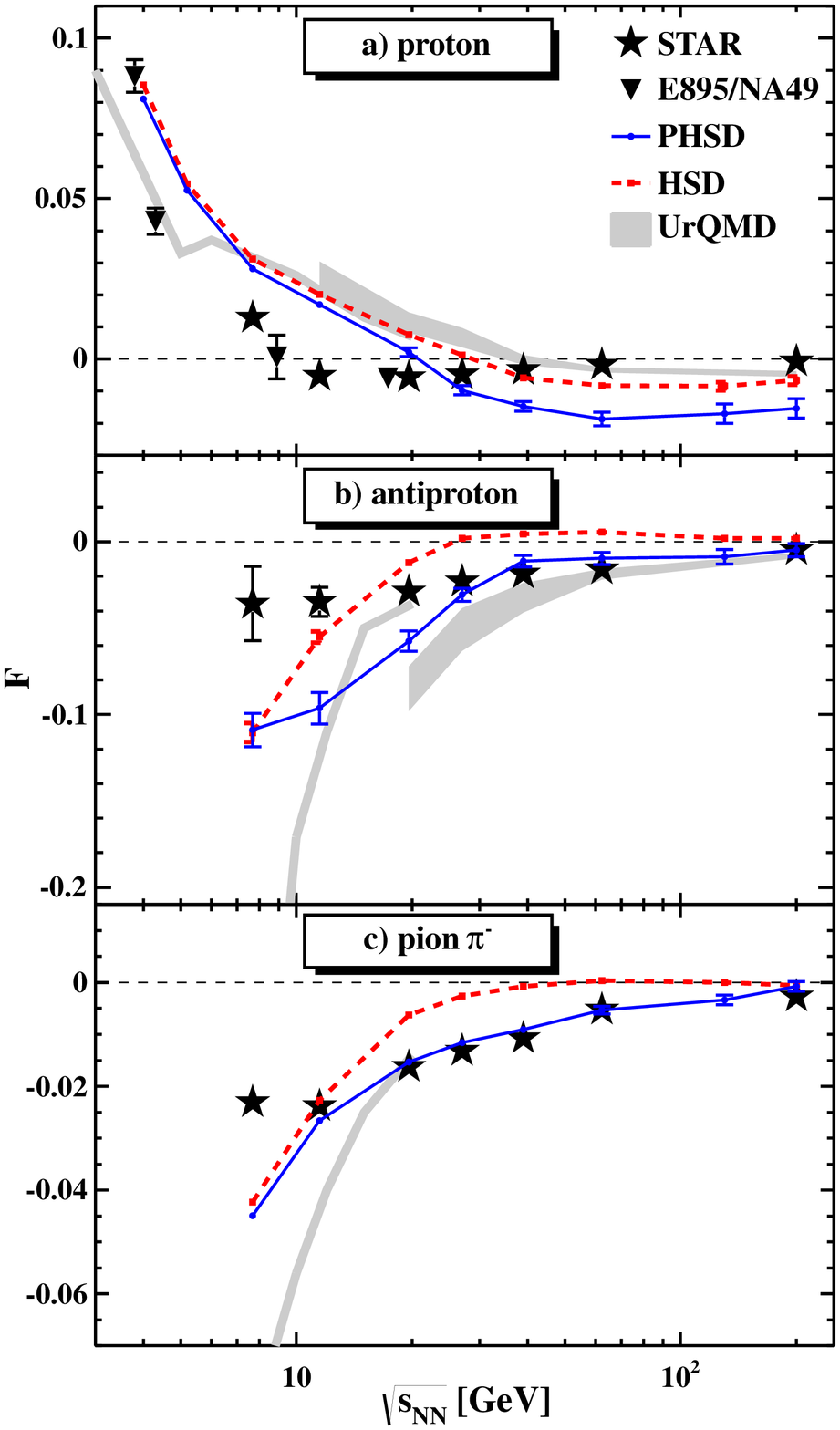}
\caption{(l.h.s.) The directed flow $v_1(y)$ for protons,
antipprotons as well as  negative pions from Au+Au collisions at
different collision energies from $\sqrt{s_{NN}}=$ 7.7 to 39 GeV {
from HSD (dashed lines) and PHSD (solid lines)}. Experimental data
are from the STAR collaboration~\cite{STAR-14}.   (r.h.s.) The beam
energy dependence of the directed flow slope $F$ at midrapidity for
protons, antiproton and charged pions from semicentral Au+Au
collisions. The shaded band corresponds to the UrQMD results as
cited in~\cite{STAR-14}. The experimental  data are from the STAR
collaboration~\cite{STAR-14} along with results of prior experiments
using comparable cuts~\cite{NA49,E895}. The PHSD/HSD calculations
have been performed without incorporating any hadronic mean fields.}
\label{fig2}
\end{figure}

The whole set of directed flow excitation functions for protons,
antiprotons and charged pions from the PHSD/HSD models is presented
in Fig. \ref{fig2} (l.h.s.) in comparison to the measured
data~\cite{STAR-14}.
The initial states in the PHSD/HSD  are simulated on an
event-by-event basis taking into account fluctuations in the
position of the initially colliding nucleons and fluctuations in the
reaction plane. This procedure is identical to that in the study of
the elliptic flow in Ref.~\cite{KBC12}. The average impact parameter
for the selected events is $b=$ 7 fm. In the simulations the
experimental acceptance $0.2\geq p_T \geq 2$ GeV/c is taken into
account  for all hadrons~\cite{STAR-14}. Note that the PHSD/HSD
calculations in Fig. 2 have been performed without incorporating any
hadronic mean fields for $\sqrt{s_{NN}} \ge$ 7.7 GeV.

At first glance, both models -- in particularly the PHSD  --
correctly reproduce the general trends in the differential $v_1(y)$
with bombarding energy: the $v_1(y)$ slope for protons is positive
at low energies ($\sqrt{s_{NN}}\le$ 20 GeV) and approaches zero with
increasing energy while antiprotons and pions have negative slopes,
respectively, in the whole energy range. In more detail: for protons
the directed flow distributions are in a reasonable agreement with
the STAR measurements in the whole range of the collision energies
considered (except for $\sqrt{s_{NN}}=$ 11.5 GeV). However, $v_1(y)$
for antiprotons agrees with the data only for the highest energies
where baryon/antibaryon pairs are dominantly produced by
hadronization. This becomes evident from a comparison to the HSD
results with $v_1(y) \approx 0$. The shape of the $v_1(y)$
distribution for antiprotons starts progressively to differ from the
measured data if we proceed from $\sqrt{s_{NN}}=$11.5 down to 7.7
GeV. In the lower energy range the HSD and PHSD results get very
close which indicates the dominance of hadronic reaction channels
(absorption and recreation). The direct flow distributions for
negative and positive pions are close to each other and also begin
to disagree with experiment in the same range of low collision
energies as for antiprotons (see Fig. \ref{fig2} (l.h.s.)). Again
the PHSD results are very close to the experimental measurements at
higher energies while the HSD results deviate more sizeably thus
stressing the role of partonic degrees of freedom in the entire
collision dynamics. The clear overestimation of the ${\bar p}$ and
$\pi^-$ slopes at $\sqrt{s_{NN}}=$7.7 GeV demonstrates that the
heavy-ion dynamics is not yet fully understood within the
string/hadron picture  at the lower energies without incorporating
any hadronic mean fields.

The characteristic slope of the $v_1(y)$ distributions at
midrapidity, $\frac{d v_1}{dy}|_{y=0}=F$, is presented in
Fig.~\ref{fig2} (r.h.s.) for all cases considered. In a first
approximation the $v_1$ flow in the  center-of-mass system may be
well fitted by  a linear function $v_1(y)=F\ y$ within the rapidity
interval $-0.5<y<0.5$. A  cubic equation is also used,
\begin{equation} \label{cub}
v_1(y)=Fy+Cy^3~ ,
\end{equation}
to obtain an estimate of the uncertainty in extracting the
coefficient $F$. The error bars in Fig.~\ref{fig2} (r.h.s.) just
stem from the different fitting procedures. Note that the energy
axis in Fig.~\ref{fig2} (r.h.s.) is extended by adding experimental
results for $\sqrt{s_{NN}}=$ 62 and 200 GeV~\cite{STAR-14}. This
representation is more delicate as compared to $v_1(y)$ in
Fig.~\ref{fig2} (l.h.s.). For protons there is a qualitative
agreement of the HSD/PHSD results with the experiment measurements:
the slope $F>0$ at low energies, however, exceeding the experimental
values by up a factor of about two; the slope crosses the line $F=$0
at $\sqrt{s_{NN}}\sim$20 GeV, which is twice larger than the
experimental crossing point, and then stays negative and almost
constant with further energy increase. However, the absolute values
of the calculated proton slopes in this high energy range are on the
level of -(0.010-0.015), while the measured ones are about -0.005.
The standard UrQMD model results, as cited in the experimental
paper~\cite{STAR-14} and in the more recent theoretical
work~\cite{SAP14}, are displayed in Fig.~\ref{fig2} (r.h.s.) by the
wide and narrow shaded areas, respectively. These results for
protons are close to those from the HSD and essentially overestimate
the slope for energies below $\sim$30 GeV but at higher energy
become negative and relatively close to the experiment. The
predictions for the pure hadronic version of the transport model HSD
(dotted lines in Fig.~\ref{fig2} (r.h.s.) slightly differ from the
PHSD results which overpredict the negative proton slope at higher
RHIC energies.

For the antiproton slopes we again observe an almost quantitative
agreement with the BES experiment~\cite{STAR-14}: with increasing
collision energy the HSD and PHSD slopes grow and then flatten above
20-30 GeV. The HSD results  saturate at $v_1(0)=0$, while the PHSD
predictions stay negative and   in good agreement with experiment
(see Fig. \ref{fig2} (r.h.s.). It is noteworthy to point out that
these PHSD predictions strongly differ from the UrQMD results which
{ no longer} describe the data for $\sqrt{s_{NN}}\le$20 GeV but are
in agreement with the measurements for higher energies. This
disagreement { might} be attributed to a neglect of the inverse
processes for antiproton annihilation~\cite{Ca02} in UrQMD as
described above.
\begin{figure}[h!]
\includegraphics[width=0.48\textwidth]{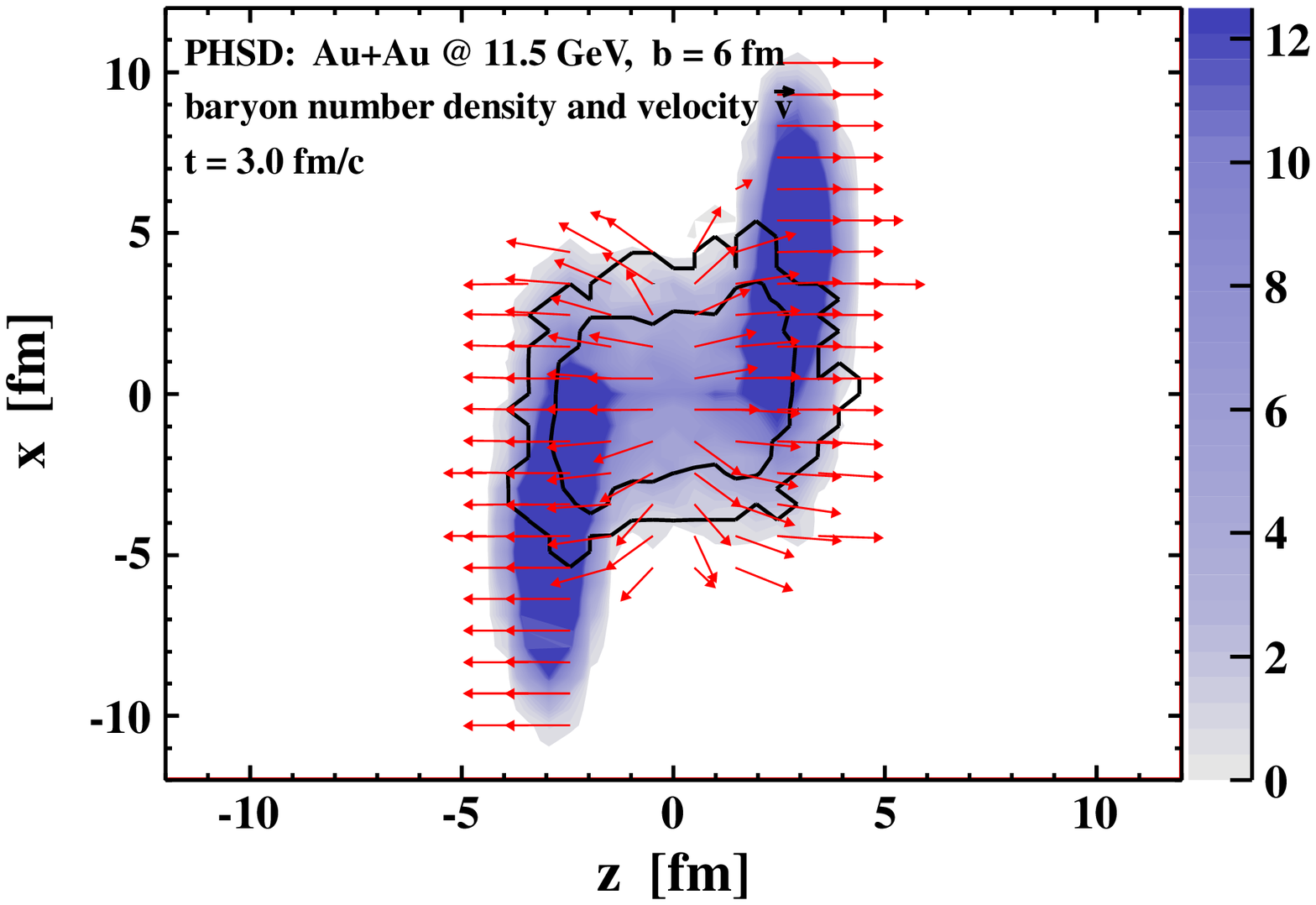}
\includegraphics[width=0.48\textwidth]{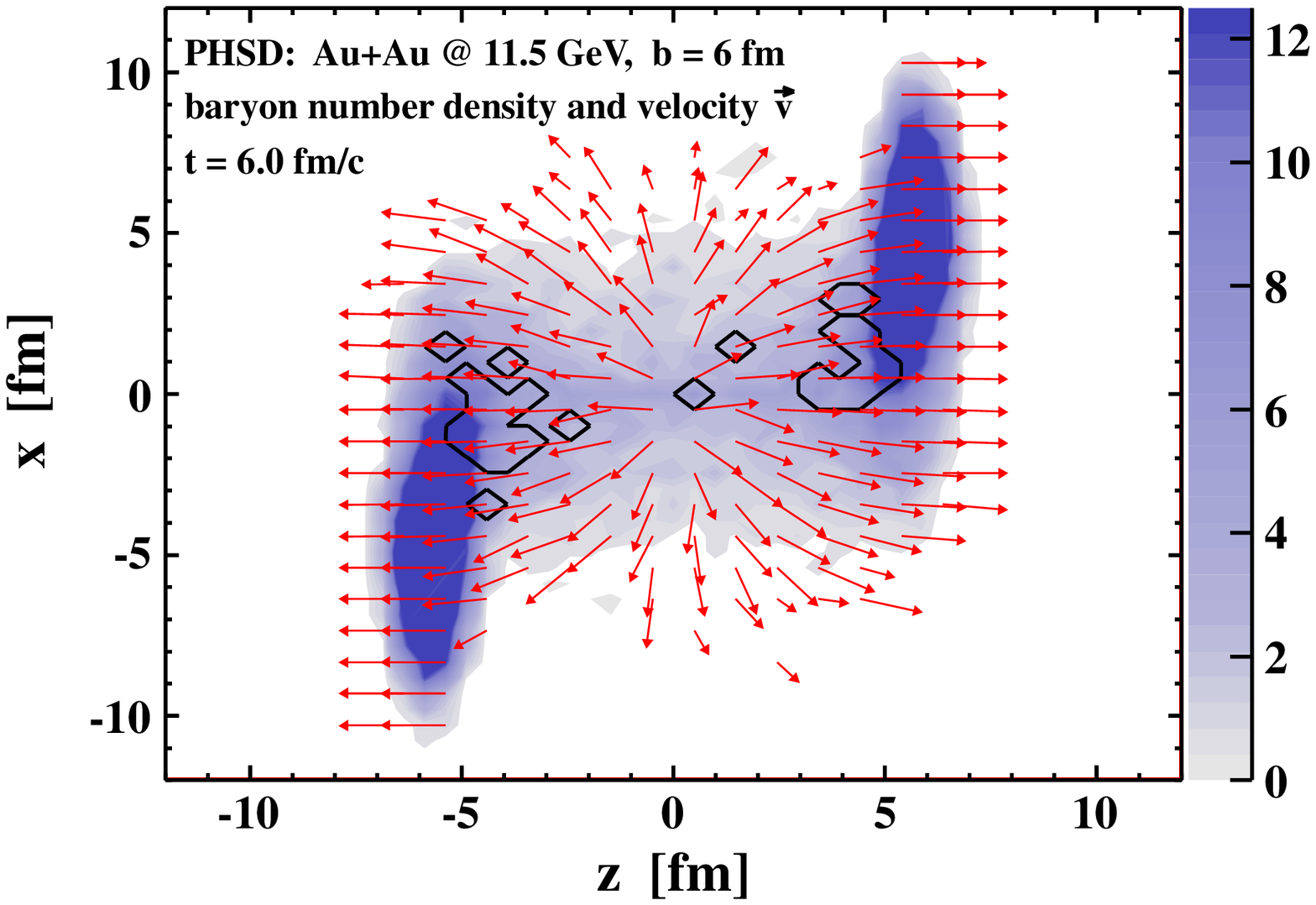}
\caption{Snapshots of the baryon energy density distribution in the PHSD model
at the time $t=3$fm/c and 6 fm/c  for Au+Au collisions and $\sqrt{s_{NN}}=$11.5 GeV.
The energy density scale is given on the right side in GeV/fm$^3$.
The solid curves display parton density levels for  0.6 and 0.01 partons/fm$^3$.
The arrows show the local velocity of baryonic matter (in relative units).
 }
\label{fig:dens}
\end{figure}

The differences between the calculations and experimental data
become apparent for the charged pion slopes  at $\sqrt{s_{NN}}\le$11
GeV: the negative minimum of the charged pion slope is deeper than
the measured one. The HSD and PHSD results practically coincide at
low energy (due to a minor impact of partonic degrees of freedom)
but dramatically differ from those of the UrQMD model for
$\sqrt{s_{NN}}\leq$20 GeV (see Fig.\ref{fig2} (r.h.s.). This
difference { might} be attributed again to a neglect of the inverse
processes for antiproton annihilation in UrQMD.

The appearance of negative $v_1$-slopes can be explained by the
evolution of the tilted ellipsoid-like shape of the participant
zone.  Snapshots of the velocity profile are shown in Fig. 3 for
times $t=$3 fm/c and 6 fm/c for  semi-peripheral Au+Au (11.5 GeV)
collisions in the background of baryon density distributions where
also  parton blobs can be identified. Indeed, among the scattered
particles there are many which move perpendicularly to the stretched
matter (antiflow) and their multiplicity increases with time.
\begin{figure}[h!]
\includegraphics[width=0.51\textwidth]{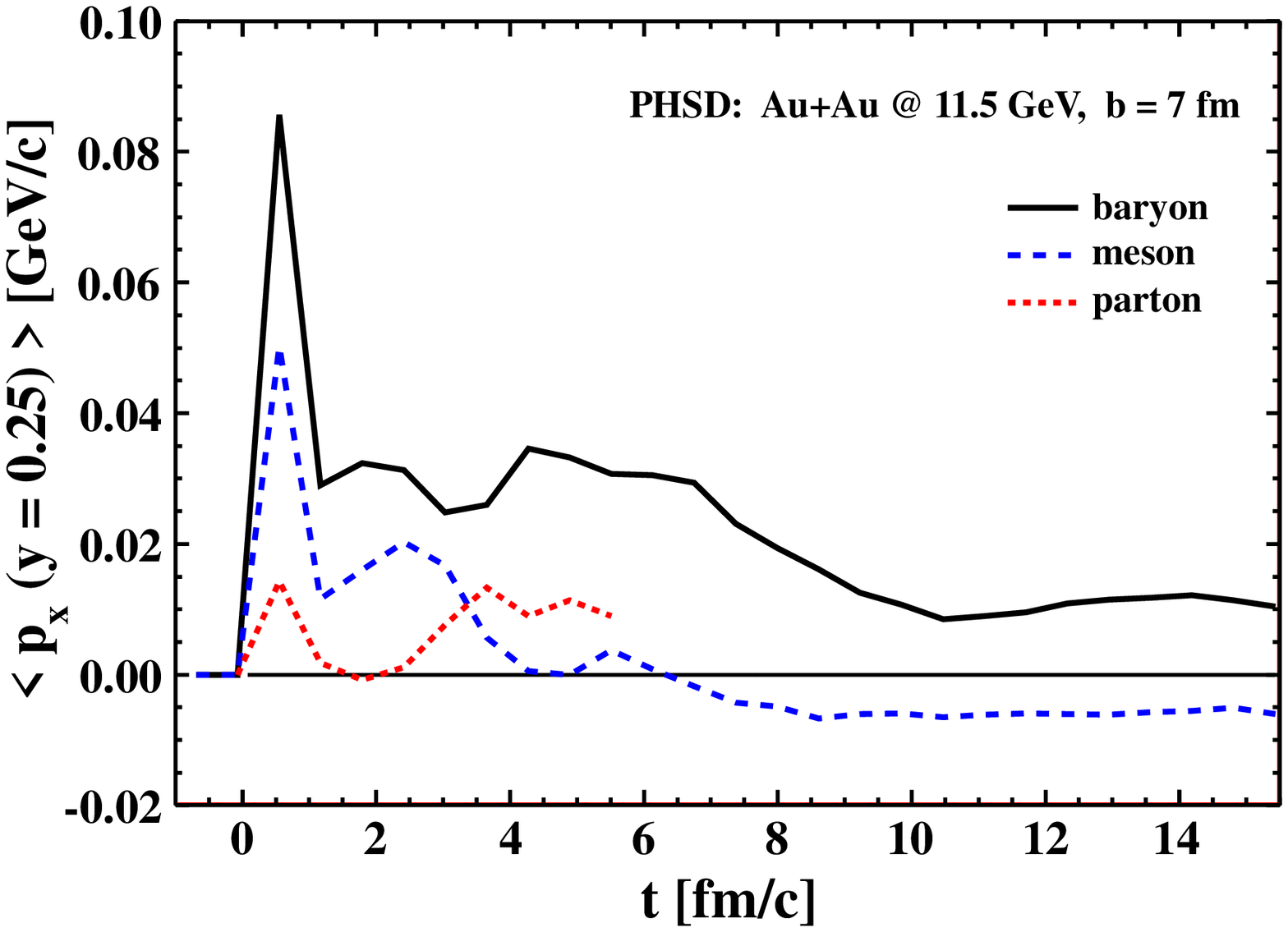}\includegraphics[width=0.48\textwidth]{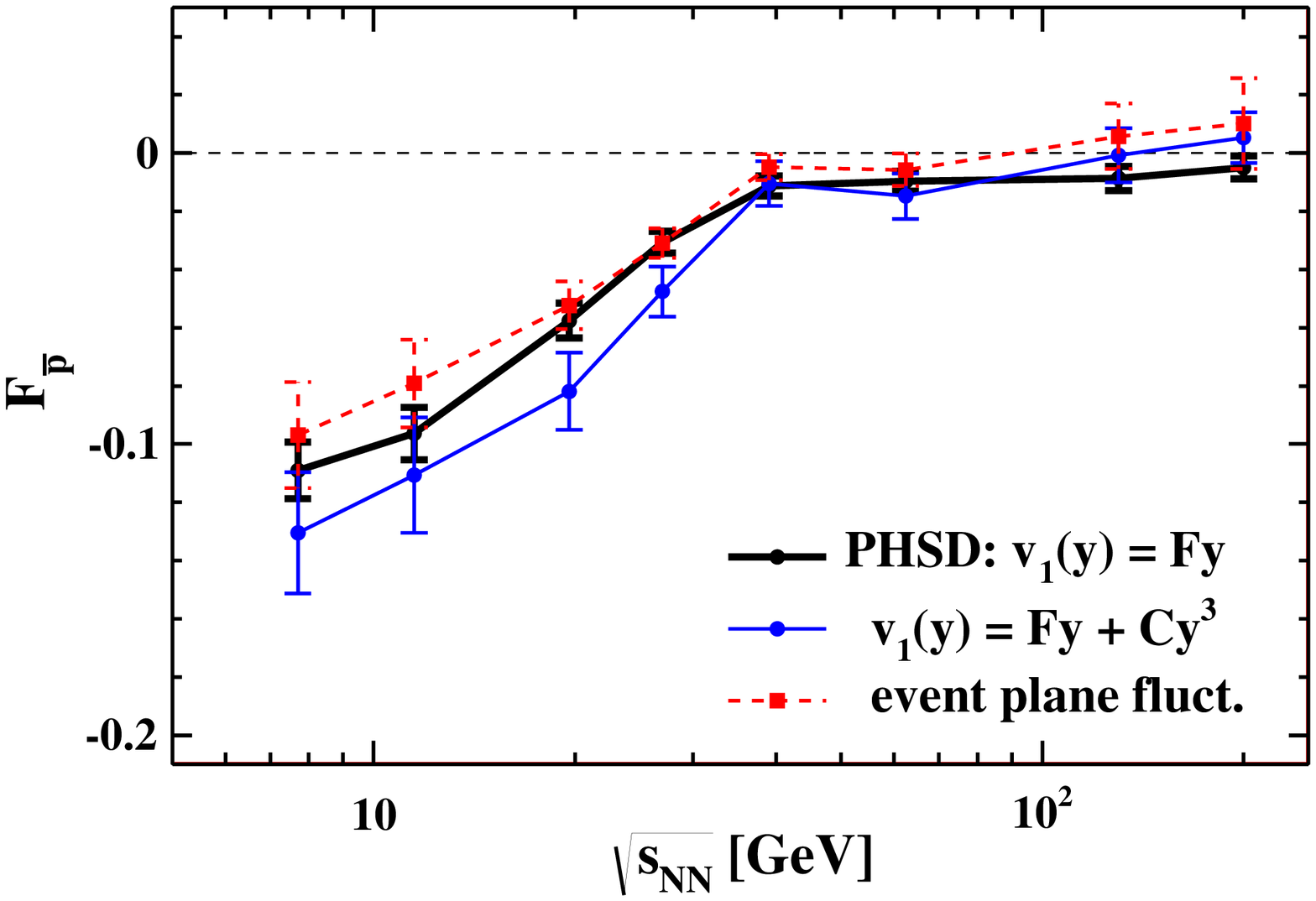}
\caption{(l.h.s.) Evolution of the average momentum projection on
the reaction plane for protons, pions and quarks at the shifted
rapidity  $y=0.25\pm$0.05. The results are given for 8.7\ 10$^4$
PHSD events  of Au+Au collisions at $\sqrt{s_{NN}}=$11.5 GeV.
(r.h.s.)  Excitation function of the antiproton slope $F_{{\bar p}}$
calculated in the PHSD model with (dotted line) and without (solid
line) including fluctuations of the reaction plane. The dotted line
corresponds to a use of the cubic equation (\ref{cub}) for the slope
calculation.} \label{fig3}
\end{figure}

However, this component is weak and it is not clear whether these
snapshots will result in observable effects for the final slope. The
solution of this question is shown in Fig.~\ref{fig3} (l.h.s.). Here
it is clearly seen that the directed flow is formed at an early
stage of the nuclear interaction. Then, the $v_1$ flow decreases
 for protons and pions reaching positive and negative slopes, respectively, in
accordance with the results in Fig.~\ref{fig2}. Thus, in agreement
with the STAR experimental data, in the considered energy range the
PHSD model predicts  for protons a smooth  $F(\sqrt{s_{NN}})$
function which is flattening at $\sqrt{s_{NN}}\geq 10$ GeV and
reveals no signatures of a possible first-order phase transition as
expected in Refs.~\cite{Ri95,Ri96,HS94,ST95,MO95,RG96,St05}. For
antiprotons the slope at midrapidity manifests a wide but shallow
negative minimum for $\sqrt{s_{NN}}\approx$30 GeV while the measured
slope is a monotonically  increasing function. It is noteworthy that
the new STAR data are consistent with the  PHSD results which
include a crossover  transition by default due to a matching of the
EoS to lattice QCD results. We note in passing that fluctuations of
the reaction plane give only a small effect on the directed flow of
hadrons which is most pronounced for antiprotons (cf. Fig.
\ref{fig3} (r.h.s.)).

\section{Directed flow of kaons and antikaons}
\label{kaon} Strange hadrons and in particulalar kaons and antikaons
provide additional information on the reaction dynamics. In
relativistic mean-field models the dispersion relation for kaons and
antikaons in the nuclear medium  can be written as
\cite{EX1,EX2,EXX,Brat2012}
\begin{equation} \label{nelson1}
\omega_{K^\pm}^2(\rho_N,{\bf p}) = \pm \frac{3}{4} \frac{\omega
}{f_K^2}\rho_N  + m_K^2 + {\bf p}^2 - \frac{\Sigma_{KN}}{f_K^2}
\rho_s  .
\end{equation}
In Eq. (\ref{nelson1}) $\Sigma_{KN}$ is the kaon-nucleon sigma term
($\approx$ 400 MeV), $m_K$ denotes the bare kaon mass, $f_K \approx$
100 MeV is the kaon decay constant, while $\rho_s$ and $\rho_N$
stand for the scalar and vector nucleon densities, respectively.
This leads to repulsive kaon mean fields $U_K > 0$ and attractive
mean fields for the antikaons $U_{\bar{K}} < 0$ at finite baryon
density \cite{EX1,EX2,Brat2012}.

\subsection{$K^\pm$ potentials}
\label{pot} 
\begin{figure}[thb]
\includegraphics[width=0.48\textwidth]{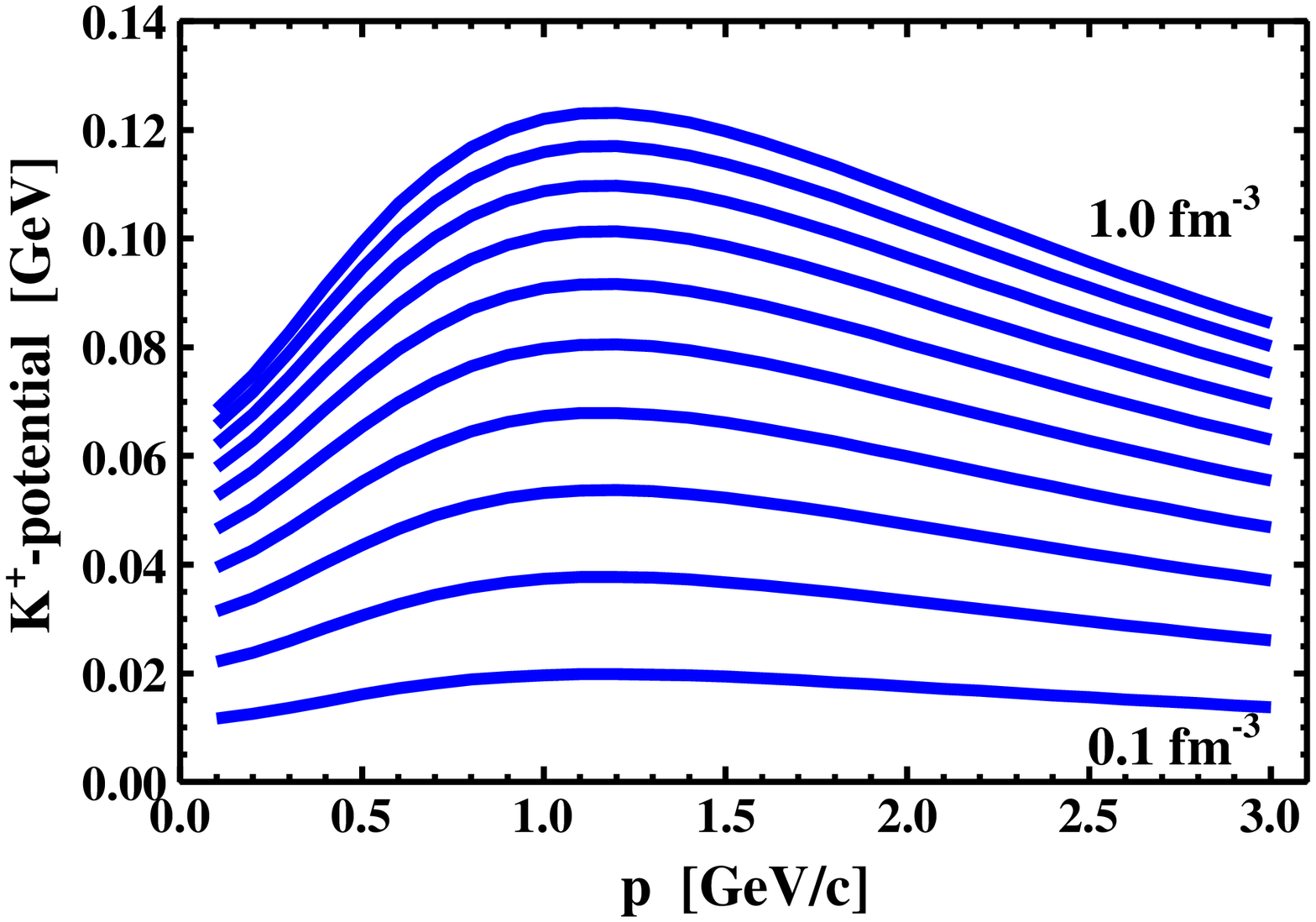} \includegraphics[width=0.48\textwidth]{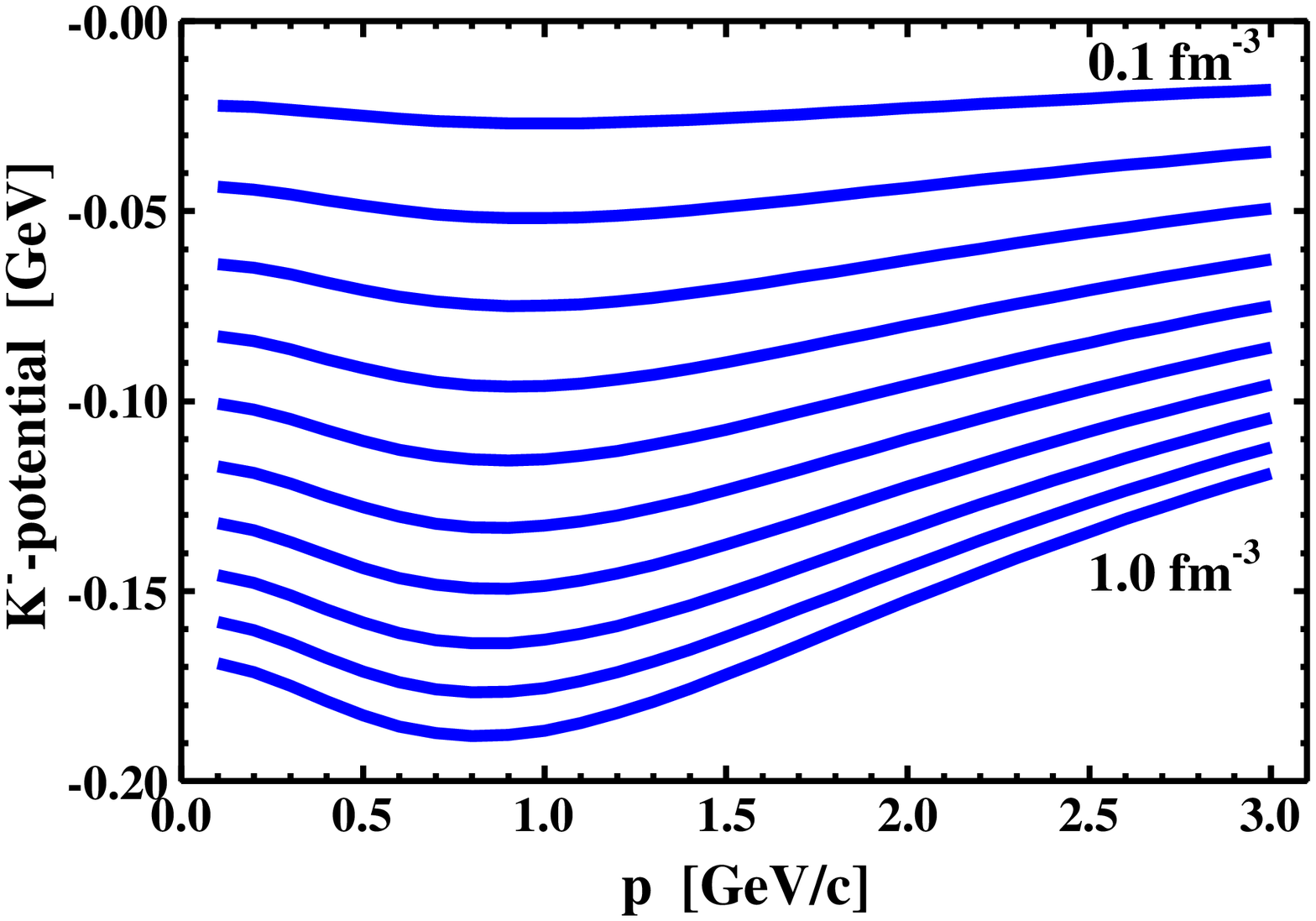}
\caption{(l.h.s.) The repulsive kaon potential as a function of the momentum $p$ with respect to the local rest frame for nuclear densities
from 0.1$fm^{-3}$ to 1.0 $fm^{-3}$ in steps of 0.1$fm^{-3}$.
(r.h.s.) Same as on the l.h.s. for the attractive antikaon potential employed.}
 \label{fig5}
\end{figure}

The microscopic calculation of kaon and antikaon
potentials (or mean fields) is more involved and can e.g. be worked
out within G-matrix theory \cite{Laura,EX4}. These calculations show
that the potentials also explicitly depend on the $K^\pm$ momentum
$p$ with respect to the local rest frame of the system. Such a
momentum-dependence can be incorporated by introducing momentum
dependent formfactors in the scalar and vector potentials and
fitting the parameters to the results from G-matrix theory up to
twice nuclear matter density. Any extrapolation to higher densities,
however, should be taken with great care since no robust information
is available so far. In order to shed some light on the possible
effects of $K^\pm$ potentials on their directed flow - especially at
FAIR/NICA energies - we incorporated the momentum-dependent
potentials for kaons and antikaons as displayed in Fig. \ref{fig5}
as a function of the momentum $p$ for nuclear densities $\rho_N$
from 0.1 - 1.0 $fm^{-3}$ in steps of 0.1 $fm^{-3}$. These
potentials, defined by
\begin{equation}
U_K({\bf p},\rho_N)=\omega_{K}(\rho_N,{\bf p})-\sqrt{{\bf p}^2 +
m_K^2},
\end{equation} typically increase/decrease with momentum and asymptotically tend
to zero again. Note that a density of 1.0 $fm^{-3}$ roughly
corresponds to 6 times nuclear matter density where the system no
longer should consist of hadronic degrees of freedom. Accordingly
these potentials are dominantly probed only up to 3 times nuclear
matter density in actual PHSD calculations since the partonic
degrees of freedom take over at higher densities and the potentials
for the strange quarks ($s, {\bar s}$) are given by the DQPM.

\subsection{Predictions for directed kaon and antikaon flows}
\label{results}
  \begin{figure}[t]
\includegraphics[width=0.48\textwidth]{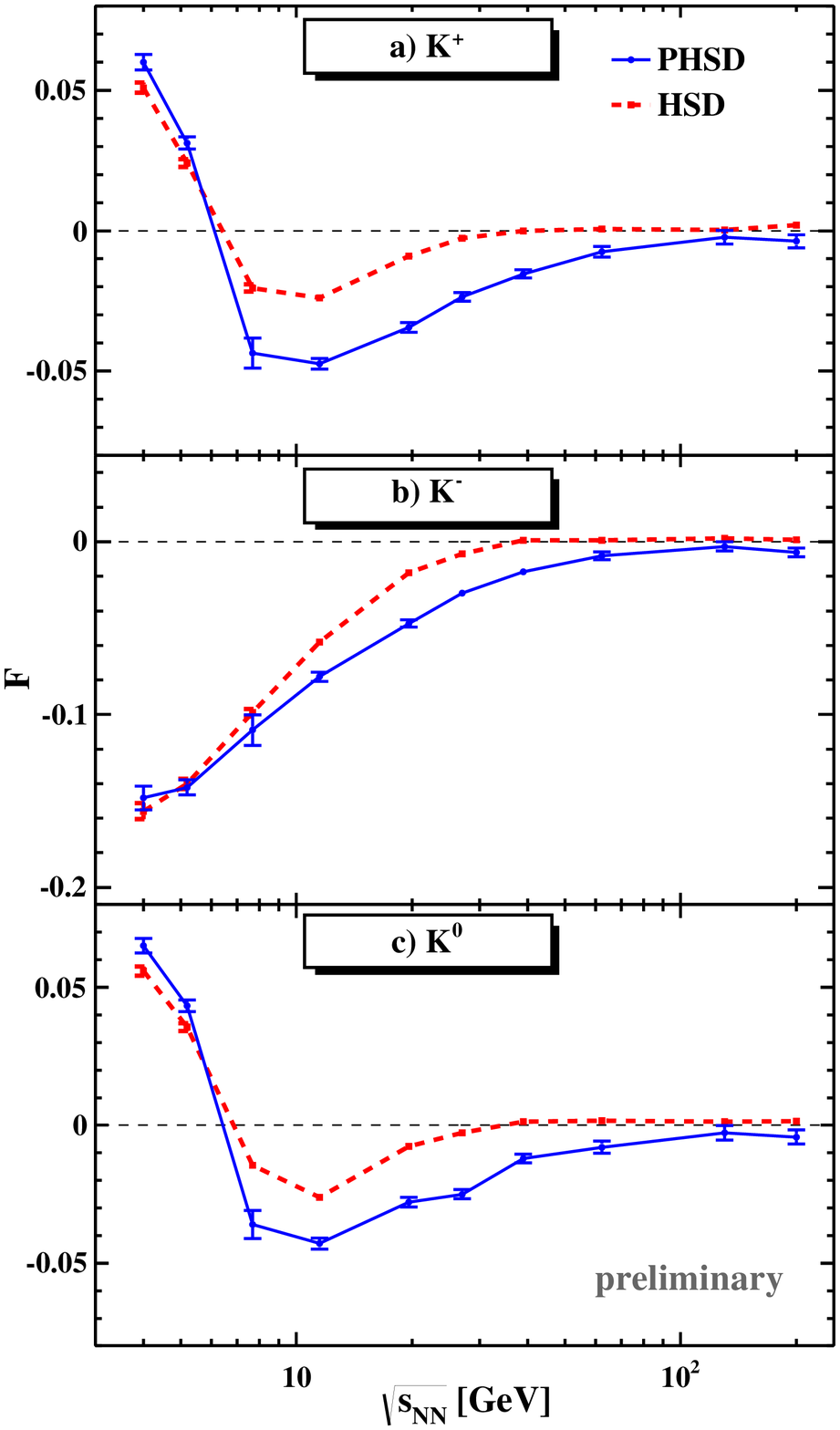} \includegraphics[width=0.48\textwidth]{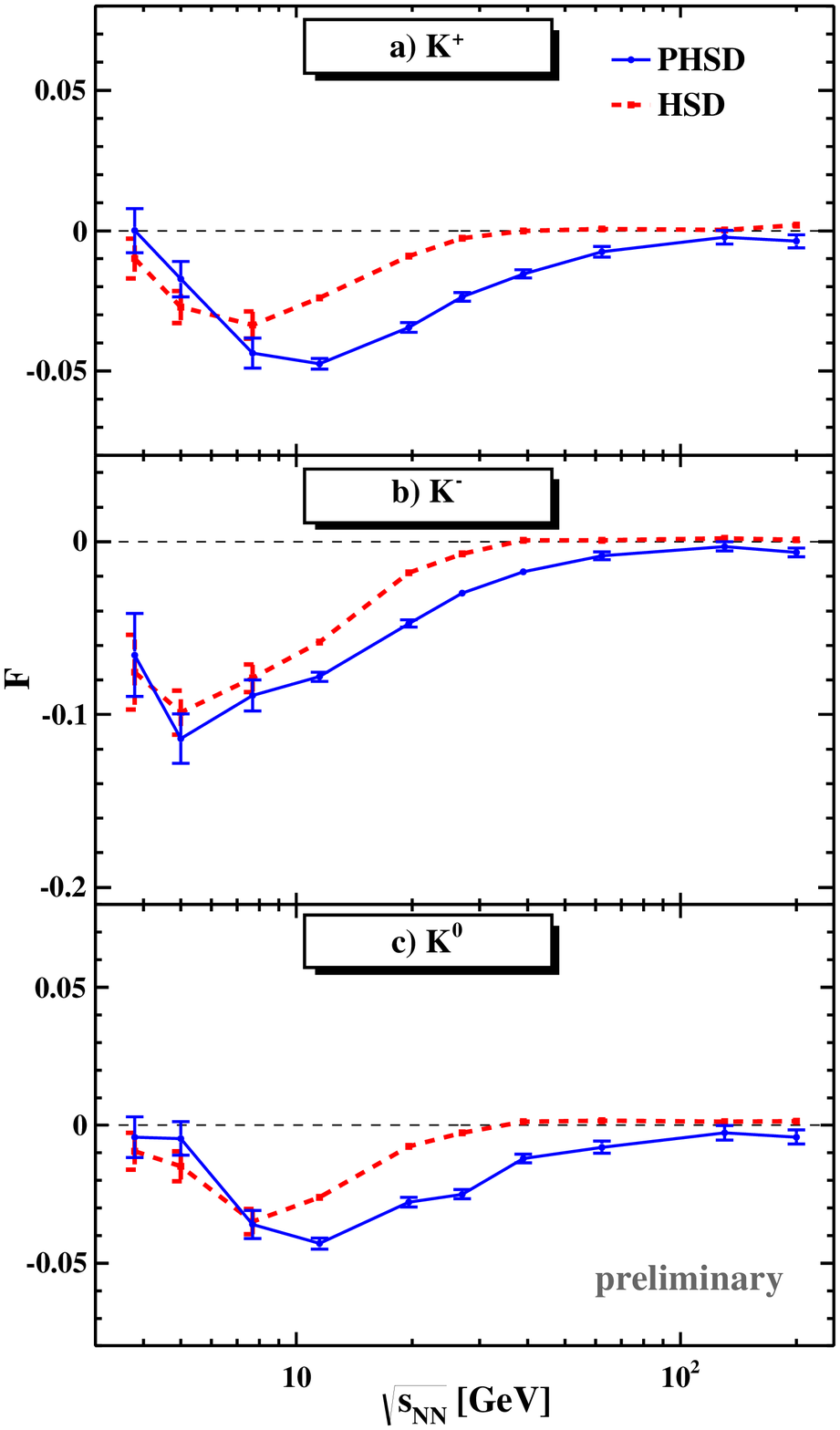}
\caption{(l.h.s.) The directed flow slope $F$ for $K^+, K^-, K^0$ as
a function of the invariant energy $\sqrt{s_{NN}}$ from lower
FAIR/NICA to RHIC energies without kaon potentials. (r.h.s.) The
directed flow slope $F$ for $K^+, K^-, K^0$ as a function of the
invariant energy $\sqrt{s_{NN}}$  with kaon potentials in the HSD
(dashed lines) and PHSD (solid lines) calculations.}
 \label{fig6}
\end{figure}
We directly step on with the preliminary results for Au+Au
collisions at b=7 fm for invariant energies of $\sqrt{s_{NN}}$ from
3 to 200 GeV, i.e. from the low FAIR/NICA energies to the top RHIC
energy. The actual results are shown in Fig. \ref{fig6} from HSD
(dashed lines) and PHSD (solid lines) for $K^+, K^-$ and $K^0$
mesons, where the lefthand column corresponds to calculations
without kaon potentials whereas the righthand column shows
calculations with the kaon potentials from Fig. \ref{fig5} included.
As in Fig. 2 (r.h.s.) we find the same pattern (without potentials)
as for pions above $\sqrt{s_{NN}} \approx$ 10 GeV. The PHSD
calculations show a larger negative flow for kaons and antikaons
than HSD; these directed flows are comparable in size with those
from the pions. The latter is due to the fact that these hadrons
dominantly emerge from  parton fusion in the hadronization process.
On the other hand, the HSD and PHSD give very similar results for
$\sqrt{s_{NN}} <$ 7 GeV which has to be attributed to the dominance
of hadronic degrees of freedom. Furthermore, we find a remarkable
sensitivity to the kaon/antikaon potentials in this low energy
(FAIR/NICA) domain when comparing the results from the left and
right columns in Fig. \ref{fig6}. The positive flow of kaons even
changes sign when including the repulsive potentials whereas the
antikaon flow is decreased in size substantially by the attractive
mean fields. Above about $\sqrt{s_{NN}}
>$ 10 GeV there is no longer a sizeable sensitivity to the
kaon/antikaon potentials. Accordingly, heavy-ion collisions at
FAIR/NICA energies have the potential to provide information on the
hadron properties (or dispersion relations) at high baryon densities
while still at moderate temperatures.

\section{Conclusions}
\label{final}

In this study the parton-hadron-string dynamics (PHSD) approach has
been applied for the analysis of the recent STAR data on the
directed flow of identified hadrons~\cite{STAR-14} in the energy
range $\sqrt{s_{NN}}=$7.7-200 GeV. The excitation functions for the
directed flows of protons, antiprotons and charged pions turn out to
be smooth functions in bombarding energy  without 'wiggle-like'
irregularities \cite{Volodya2014} as expected before in Refs.~\cite{
Ri95,Ri96,HS94,ST95,MO95,RG96,St05}. Our results differ from the
standard UrQMD model at lower bombarding energies as included in
Ref.~\cite{STAR-14} and the recent theoretical analysis in
Ref.~\cite{SAP14}. The microscopic PHSD transport approach
reproduces the general trend in the differential $v_1(y)$ excitation
function  and leads to an almost quantitative agreement for protons,
antiprotons and pions especially at higher energies. We attribute
this success to the Kadanoff-Baym dynamics incorporated in PHSD
(with more accurate spectral functions) as compared to a
Boltzmann-like on-shell transport model (UrQMD) and the account for
parton dynamics  also in this 'moderate' energy range. The latter is
implemented in PHSD in line with the equation of state from lattice
QCD \cite{Wuppertal}. The formation of the parton-hadron mixed phase
softens the effective EoS in PHSD and describes a crossover
transition (in line with the lattice QCD EoS). Accordingly, the PHSD
results differ from those of HSD where no partonic degrees of
freedom are incorporated. A comparison of both microscopic models
has provided detailed information on the effect of parton dynamics
on the directed flow (cf. Fig. \ref{fig2}).

Antiprotons have been shown to be particularly interesting. In
HSD/PHSD we include antiproton annihilation into several mesons
while taking into account also  the inverse processes of $p\bar{p}$
creation in multi-meson interactions by detailed
balance~\cite{Ca02}. Related kinetic models (including  UrQMD) which
neglect the inverse processes for antiproton annihilation at lower
energies { do not} describe the data on the directed flow of hadrons
$v_1(y)$. We note in passing that 3FD  hydrodynamics provides the
best results with a crossover EoS for the quark-hadron phase
transition \cite{Volodya2014} which by default is implemented in
PHSD.

Still sizeable discrepancies with experimental measurements in the
directed flow characteristics are found for the microscopic kinetic
models at $\sqrt{s_{NN}}\leq$ 15 GeV and are  common  for both HSD
and PHSD (and UrQMD \cite{Brat04}) since the partonic degrees of
freedom are subleading at these energies. We recall that the flow
observables are not the only ones where the kinetic approaches have
a problem in this energy range. Another long-standing issue is the
overestimation of pion production as seen in Fig.~\ref{fig1} in the
energy regime around the 'horn' in the $K^+/\pi^+$ meson
ratio~\cite{GG99,CBJ00} which before has been  related to a
first-order phase transition or to the onset of
deconfinement~\cite{GGS11}. Our flow analysis shows no indication of
a first-order transition in the energy range investigated. However,
we have found further strong evidence that the dynamics of heavy-ion
reactions at lower SPS or FAIR/NICA energies is far from being
understood especially on the hadronic level without including
hadronic mean fields (potentials).

On the other hand, we could demonstrate that kaon and antikaon
potentials have a large impact on the kaon and antikaon flows for
$\sqrt{s_{NN}}\leq$ 7 GeV where the hadronic dynamics dominate.
Presently, these potentials are not well known at high baryon
densities and large momenta which provides the experimental
perspective to shed further light on this issue. Furthermore, also
baryon and antibaryon potentials will have an impact on the hadronic
flow patterns as demonstrated in Refs. \cite{Danil,Sahu,JEM}. Note
that the latter mean fields have been discarded in our present
studies. We speculate that extended theoretical approaches including
consistently chiral partners as well as a restoration of chiral
symmetry at high baryon density and/or temperature might lead to a
solution of the current problems as well as precise experimental
studies at FAIR, NICA or within the BES II program at
RHIC~\cite{CBMbook}.

\vspace{0.5cm}

{\bf Acknowledgments} \\ The authors are thankful to Yu. B. Ivanov
for illuminating discussions and valuable suggestions as well as for
related calculations within the 3FD approach.


\begin{thebibliography}{99}
%
\bibitem{VPS10}S. A. Voloshin, A. M. Poskanzer and R. Snellings,
in Landolt-Boernstein New Series, I/23,  p. 5-54, edited by R. Stock
(Springer-Verlag, 2010).
%
\bibitem{So10} P. Sorensen,
In Quark-Gluon Plasma 4, ed. by R. Hwa and X.N. Wang, World
Scientific (2010).
%
\bibitem{PV98} A.M. Poskanzer and S.A. Voloshin,
Phys. Rev. C {\textbf{58}}, 1671 (1998).
%
\bibitem{STAR10} STAR Collaboration,
arXiv:1007.2613.
\bibitem{Danil} P. Danielewicz, Nucl. Phys. A {\textbf{673}}, 375 (2000).

\bibitem{Sahu} P. K. Sahu and W. Cassing, Nucl. Phys. A {\textbf{712}},
357 (2002).

\bibitem{JEM} M. Isse, A. Ohnishi {\it et al.}, Phys. Rev. C
{\textbf{72}}, 064908 (2005).
%
\bibitem{CBMbook} P. Senger {\it  et al.}, Lect. Notes Phys. {\textbf{814}}, 681 (2011).
%
\bibitem{So97} H. Sorge,
Phys. Rev. Lett. {\textbf{78}}, 2309 (1997).
%
\bibitem{HWW99} N. Herrmann, J. P. Wessels, and T. Wienold, Ann. Rev. Nucl. Part. Sci.
{\textbf{49}}, 581 (1999).
%
\bibitem{SH92} E. Schnedermann and U. Heinz,
Phys. Rev. Lett. {\textbf{69}}, 2908 (1992).
%
\bibitem{KKP95} D. E. Kahana, D. Keane, Y. Pang, T. Schlagel and S. Wang,
Phys. Rev. Lett. {\textbf{74}}, 4404 (1995).
%
\bibitem{E877} J. Barrette {\it et al.} (E877 Collaboration),
Phys. Rev. Lett. {\textbf{73}}, 2532 (1994).
%
\bibitem{NA44} I. G. Bearden {\it et al.} (NA44 Collaboration),
Phys. Rev. Lett. {\textbf{78}}, 2080 (1997).
%
%
\bibitem{NA49} C. Alt {\it et al.} (NA49 Collaboration),
Phys. Rev. C {\textbf{68}}, 034903 (2003).
%
\bibitem{E877-7}E877 Collaboration, J. Barrette
{\it et al.}, Phys. Rev. C {\textbf{55}}, 1420 (1997); J. Barrette
{\it et al.}, Phys. Rev. C {\textbf{56}}, 3254 (1997).
%
\bibitem{RR97} W. Reisdorf and H.G. Ritter, Annu. Rev. Nucl. Part. Sci. {\textbf{47}}, 663 (1997).
%
\bibitem{WA98} WA98 Collaboration, M.M. Aggarwal {\it et al.},
nucl-ex/9807004.
%
\bibitem{INN00} Yu. B. Ivanov {\it et al.},
Acta Phys. Hung. New Ser. Heavy Ion Phys. {\textbf{15}}, 117 (2002).
%
\bibitem{Ri95}  D. H. Rischke {\it et al.},
 Heavy Ion Phys. {\textbf{1}}, 309 (1995).
%
\bibitem{Ri96} D. H. Rischke,
Nuclear Physics A {\textbf{610}},  88 (1996).
%
\bibitem{HS94}C. M. Hung and E. V. Shuryak,
 Phys. Rev. Lett. {\textbf{75}}, 4003 (1995).
 %
\bibitem{ST95} A.A. Shanenko and V.D. Toneev, JINR Rap. Com.5[73], 21
(1995); E.G. Nikonov, A.A. Shanenko and V.D. Toneev, Heavy Ion Phys.
{\textbf{4}}, 333 (1996).
%
\bibitem{MO95} L. Mornas and U. Ornik. Nucl. Phys. A {\textbf{587}}, 828 (1995).
%
\bibitem{RG96} D. Rischke and M. Guylassy. Nucl. Phys. A {\textbf{597}}, 4 (1996).
%
\bibitem{St05} H. St\"ocker,
Nucl. Phys. A {\textbf{750}}, 121 (2005).
%


\bibitem{STAR-14} STAR Collaboration: L. Adamczyk, {\it et al.},
Phys. Rev. Lett. {\textbf{112}}, 162301 (2014).
%
\bibitem{SAP14} J. Steinheimer {\it et al.},
Phys. Rev. C {\textbf{89}}, 054913 (2014).
%
\bibitem{Volodya2014} V. Konchakovski, W. Cassing and V. D. Toneev,
Phys. Rev. C {\textbf{90}}, 014903 (2014)


\bibitem{JCG04} S. Juchem, W. Cassing, and C. Greiner, Phys. Rev. D
{\textbf{69}}, 025006 (2004);  Nucl. Phys. A {\textbf{743}}, 92
(2004).

%
\bibitem{CB09}
W. Cassing, E. L. Bratkovskaya, Nucl. Phys. A {\textbf{831}}, 215
(2009); Phys. Rev. C {\textbf{78}}, 034919 (2008); W. Cassing, Nucl.
Phys. A {\textbf{791}}, 365 (2007).

%
\bibitem{Ca09} W. Cassing,  E. Phys. J. ST {\textbf{168}}, 3 (2009).
%
\bibitem{KBC12} V. P. Konchakovski {\it et al.},
Phys. Rev. C {\textbf{85}}, 011902 (2012).
%
\bibitem{Linnyk2011}
O. Linnyk {\it et al.}, Phys. Rev. C {\textbf{84}} (2011) 054917;
Phys. Rev. C {\textbf{85}} (2012) 024910; Phys. Rev. C {\textbf{87}}
(2013) 014905.

\bibitem{To12} V. D. Toneev {\it et al.},
Phys. Rev. C {\textbf{85}}, 034910 (2012).
%

\bibitem{EC96} W. Ehehalt and W. Cassing, Nucl. Phys. A {\textbf{602}}, 449 (1996).
%
\bibitem{PhysRep} W. Cassing and E. L. Bratkovskaya,
Phys. Rep. {\textbf{308}}, 65 (1999).
%
\bibitem{CBJ00} W. Cassing, E. L. Bratkovskaya, S. Juchem,
Nucl. Phys. A {\textbf{674}}, 249 (2000).
%
\bibitem{Brat04} E. L. Bratkovskaya {\it et al.},
Phys. Rev. C {\textbf{69}}, 054907 (2004).

\bibitem{ABS06} A. Andronic, P. Braun-Munzinger and J. Stachel, Nucl.
Phys. A {\textbf{772}}, 167 (2006).
%
\bibitem{Zhu12} X. Zhu [STAR Collaboration], Acta Phys. Polon. Supp.
{\textbf{5}}, 213 (2012).
%
\bibitem{BGL} C. Blume {\it et al.},
https://edms.cern.ch/document/1075059
%
\bibitem{BM11} C. Blume and C. Markert, Prog. Part. Nucl. Phys. {\textbf{66}},
834 (2011).
%
\bibitem{Fritiof}  {  H.~Pi, Comp.~Phys.~Commun.~{\textbf{71}}, 173
(1992); T.~Sj\"ostrand {\it et al.},
Comp.~Phys.~Commun.~{\textbf{135}}, 238 (2001).}
%
\bibitem{Ca02} W. Cassing,
Nucl. Phys. A {\textbf{700}}, 618 (2002).
%
\bibitem{Bass}
    S.A.~Bass {\it et al.},
    Prog. Part. Nucl. Phys. {\textbf{42}}, 279 (1998).
%
\bibitem{Br99} L.V. Bravina {\it et al.}, J. Phys. G {\textbf{25}}, 351 (1999);
Phys. Rev. C {\textbf{62}}, 064906 (2000).
%
\bibitem{LCLM01} A.B. Larionov, W. Cassing, S. Leopold, and U. Mosel, Nucl. Phys. A {\textbf{696}},
619 (2001).
%
\bibitem{JSSSG94} A. Jahns, C. Spieles, H. Sorge, Horst St\"ocker, and W. Greiner,
Phys. Rev. Lett. {\textbf{72}}, 3464 (1994).

\bibitem{E895} H. Liu {\it et al.} (E895 Collaboration), Phys. Rev. Lett. {\textbf{84}}, 5488 (2000).

\bibitem{E877-00} J. Barrette {\it et al.} (E877 Collaboration),
Phys. Lett. B {\textbf{485}}, 319 (2000).





\bibitem{Wuppertal}  Y. Aoki {\it et al.}, Phys. Lett. B \textbf{643}, 46 (2006); S. Borsanyi {\it et al.},
JHEP \textbf{1009}, 073 (2010); JHEP {\textbf{1011}}, 077 (2010);
JHEP {\textbf{1208}}, 126 (2012); Phys. Lett. B {\textbf{730}}, 99
(2014).

\bibitem{EX1} G.-Q. Li, C. M. Ko, and X. S. Fang, Phys. Lett. B
{\textbf{329}}, 149 (1994).
\bibitem{EX2} J. Schaffner, A. Gal, I. N. Mishustin, H. St\"ocker,
and W. Greiner, Phys. Lett. B {\textbf{334}}, 268 (1994).

\bibitem{EXX}
E.L. Bratkovskaya, W. Cassing, and U. Mosel,  Nucl. Phys. A
\textbf{622}, 593 (1997).

\bibitem{Brat2012}
H. Oeschler, Y. Leifels, E. L. Bratkovskaya, and J. Aichelin, Phys.
Rept. {\textbf{510}}, 119 (2012).


\bibitem{Laura} L. Tol\'os, A.  Ramos, and A. Polls, Phys. Rev. C
{\textbf{65}}, 054907   (2002).


\bibitem{EX4} W. Cassing, L. Tol\'os, E. L. Bratkovskaya, and A. Ramos,
Nucl. Phys. A {\textbf{727}}, 59 (2003).
%
\bibitem{GG99} M. Gazdzicki and M. I. Gorenstein, Acta Phys. Polon.
B {\textbf{30}}, 2705 (1999).

\bibitem{GGS11} M. Gazdzicki, M. Gorenstein and P. Seyboth,
Acta Phys. Polon. B {\textbf{42}}, 307 (2011).

\end{thebibliography}
\end{document}